\documentclass{optica-article}

\journal{opticajournal} % for journals or Optica Open

\articletype{Research Article}

\usepackage{lineno}
\usepackage{gergo-config}   % GN: contains all includes and custom commands I use (see gergo-config.sty)

\newcommand{\CEP}{\mathrm{CEP}} % CEP in math mode
\newcommand{\MSE}{\mathrm{MSE}} % MSE in math mode
\newcommand{\MSEper}{\mathrm{MSE\%}} % MSE% in math mode
\newcommand{\Ntrain}{N_\mathrm{train}} % N_train
\newcommand{\Ntest}{N_\mathrm{test}} % N_test

\begin{document}

\title{MIR laser CEP estimation using machine learning concepts in bulk high harmonic generation}

\author{Bal\'azs Nagyill\'es\authormark{1,2, $\dag$}, Gergely N. Nagy\authormark{1,2,$\dag$}, B\'alint Kiss\authormark{1}, Eric Cormier\authormark{1,3}, P\'eter F\"oldi\authormark{1,2}, Katalin Varjú\authormark{1,2}, Subhendu Kahaly\authormark{1}, Mousumi Upadhyay Kahaly\authormark{1,2,*} and Zsolt Diveki\authormark{1,**}}

\address{\authormark{1}ELI ALPS, ELI-HU Non-Profit Ltd., Wolfgang Sandner utca 3., Szeged 6728, Hungary\\
\authormark{2}Institute of Physics, University of Szeged, D\'om t\'er 9, H-6720 Szeged, Hungary\\
\authormark{3}Laboratoire Photonique Num\'erique et Nanosciences (LP2N), UMR 5298, CNRS-IOGS-Universit\'e Bordeaux, 33400 Talence, France\\
\authormark{$\dag$}Authors contributed equally to this work}

% \authormark{3}Currently with the Department of Electronic Journals, Optica Publishing Group, 2010 Massachusetts Avenue NW, Washington, DC 20036, USA}

\email{\authormark{*}mousumi.upadhyaykahaly@eli-alps.hu}
\email{\authormark{**}zsolt.diveki@eli-alps.hu}
%% email address is required; see note below about the corresponding author designation

% use {asbstract*} to suppress the copyright line. Copyright information will be added in production

\begin{abstract*}
Monitoring the carrier-envelope phase (CEP) is of paramount importance for experiments involving few cycle intense laser fields. Common measurement techniques include f-2f interferometry or stereo-ATI setups. These approaches are adequate, but are challenging to implement on demand, at different locations as additional metrology tools, in intense few cycle laser-matter interaction experiments, such as those prevalent in sophisticated user beamlines. In addition there are inherent difficulties for CEP measured at non-conventional laser wavelengths (like e.g. mid infrared) and measurements above 10 kHz laser repetition rates, on single shot basis. Here we demonstrate both by simulations and by experiments a machine learning (ML) driven method for CEP estimation in the mid infrared, which is readily generalizable for any laser wavelength and possibly up to MHz repetition rates. The concept relies on the observation of the spectrum of high harmonic generation (HHG) in bulk material and the use of ML techniques to estimate the CEP of the laser. Once the ML model is trained, the method provides a way for cheap and compact real-time CEP tagging. This technique can complement the otherwise sophisticated monitoring of CEP, and is able to capture the complex correlation between the CEP and the observable HHG spectra.

\end{abstract*}

%%%%%%%%%%%%%%%%%%%%%%%%%%  body  %%%%%%%%%%%%%%%%%%%%%%%%%%
\section{Introduction}

High harmonic generation (HHG) relies on the highly non-linear interaction between ultrashort intense laser pulses and matter \cite{Nayak2019}, and has been demonstrated with a wide range of driving laser wavelengths \cite{Popmintchev2010,Marceau2017,Johnson2018}. Mid-infrared (MIR) driving lasers in particular have two appealing features. On the one hand, because of the wavelength scaling of the ponderomotive energy \cite{Popmintchev2010} $U_p \propto I_L\lambda^2$, the cut-off energy of the generated harmonics in gas can be extended to the keV spectral regime \cite{ Popmintchev2010,Johnson2018,Fu2020}, providing X-rays with unmatched temporal and spatial qualities. On the other hand, because of its low linear excitation rate, ultrashort MIR lasers can perform HHG in transparent solid state media \cite{Ghimire2010, Park2022}, like semiconductors or dielectrics, well below the damage threshold of the material \cite{Simanovskii2003,Migal2020}. 

In the few cycle regime, the carrier-envelope phase (CEP) of an ultrashort laser, which indicates the phase of the carrier wave with respect to the peak of the intensity envelope of the pulse, has significant impact on the strong field interaction and the HHG process. Therefore accurate measurement/monitoring and proper control of the CEP is of paramount importance, and is essential for multiple  applications in attoscience relevant to chemistry \cite{Kbel2016}, atomic and molecular physics \cite{Hanus2020} and lightwave electronics \cite{Krger2011}, to mention a few.

There are different ways to measure the CEP of few cycle pulses, utilizing different types of interactions, for example, like observation of half cycle cut off in HHG spectra from gas \cite{Haworth2006} or quantum interference in semiconductors \cite{Roos2005, Fortier2004}. Nontheless, two other techniques have become predominant. The first, one uses f-2f interferometry where a large bandwidth fundamental spectrum overlaps with its second harmonic signal and the appearance of the spectral fringes reveals the relative CEP of the laser \cite{Xu1996}. Single-shot CEP measurement above 10 kHz is very challenging with f-2f, although several methods allows one to reach MHz repetition rates\cite{Feng2013, Kurucz2019,Guo:23} and the overlap between the fundamental and the second harmonic's spectral region might lie outside some common detectors. The second, technique is based on the measurement of stereographic above threshold ionization (Stereo-ATI) signal \cite{Wittmann2009}, which is capable of performing single-shot absolute measurements at high repetition rate \cite{Hoff2018}. But, in spite of its advantages, Stereo-ATI needs sophisticated and expensive instrumentation and is an in-vacuum CEP metrology tool. The last point makes it difficult to integrate into and permanently keep in place as a metrology tool inside the sophisticated existing high repetition rate attosecond beamlines like those existing at ELI-ALPS \cite{Shirozhan2024}.

Recently new techniques relying on solid HHG were proposed to measure the (relative) CEP of the laser \cite{Hollinger2020} where the overlap between adjacent harmonic orders was exploited. The technique could be scaled to longer driving wavelengths, however the appearance of the interference pattern between adjacent harmonics is not a general condition. If the latter condition is not fullfilled an alternative approach was proposed providing the CEP stability shot-to-shot \cite{Leblanc:20} or time domain based electric field reconstruction in solids using a delayed replica of the driving laser to perturb the HHG process \cite{Awad2024}. 

During the last decade the applications of machine learning techniques have slowly entered not just into the every day life but into various fields of science, too. At XFELs \cite{Sanchez-Gonzalez2017} it was successfully applied to improve and accelerate the metrology of the emitted radiation. Convolutional neural networks have been applied to reconstruct the temporal shape of femtosecond laser pulses \cite{Toth2023}. Interestingly, their model was robust enough to retrieve the spectral amplitude and phase from experimental second harmonic generation-frequency resolved optical gating spectrograms while being trained only with simulated spectrograms. Recently, independently from our study, there was a theoretical proposition \cite{Klimkin2023} to reconstruct the band structure of a crystal and at the same time characterize the driving few-cycle laser pulse in the solid HHG process, including both their chirp and CEP, relying on the training of deep neural network models. In another theoretical study \cite{Yang2022}, the CEP dependence of the solid HHG was combined with deep learning models to retrieve the band structure of MgO crystal.

In this report, we combine the benefits of solid HHG and machine learning in order to propose a concept that helps tagging the driving laser's relative CEP in a simple setup that is instrumentally not demanding while still offering high repetition rate tagging. First, we demonstrate the feasibility of such tagging relying on simulated harmonic spectra from thin ZnO crystal based on a simple 1D model for bulk ZnO \cite{tsatrafyllis2019quantum}, that correctly reproduces its semiconducting features, while proving useful for accurate retrieval of the spectral phases of the incident few-cycle pulse, despite some amplitude noise and phase jitter.
Then we show that even without large training data the machine learning model can still achieve good relative CEP estimations. Next, relying on our estimations on the number of needed training data we experimentally demonstrate that solid HHG spectra is indeed a good indicator of the relative CEP of the driving laser.

\section{Formulation of our approach}

In most of the ultrafast light matter interactions, including solid HHG as well, the temporal/spectral profile of the driving laser exhibit a deterministic influence on the outcome. In order to enable exploitation of this feature and utilize it for laser CEP estimations, two prerequisites need to be fulfilled. Firstly, there should be a direct, but not necessarily obvious or even explicit, correlation between the harmonic spectrum and the laser CEP in solid HHG. Secondly, this correlation should manifest as a one-to-one mapping, ensuring that the high harmonic spectrum of a CEP scan exhibits a periodicity of 2$\pi$, otherwise CEP prediction would be limited for only on a fraction of the full range.

In the case of few cycle lasers, where the amplitude of the electric field changes substantially from one half cycle to the other, interband harmonics are generated at different times in each cycle \cite{Ghimire2018, You2017}. This causes a phase shift (attochirp) between the same harmonics originating from different half cycles. By changing the CEP of the laser, the timing of the emission of given harmonics can be directly controlled, causing the emergence of CEP dependent patterns in the harmonic spectrum \cite{Schubert2014, You2017, Ghimire2018}. Therefore, the first requirement, the precense of a CEP-dependent feature in the spectrum, is fulfiled. 
While CEP dependence of harmonic generation though intraband processes in mid-infrared regime is claimed to be insignificant \cite{You2017NatComm, Garg2018}, with increasing driving wavelength, signatures of CEP dependence of intraband harmonics become more prominent \cite{Schubert2014, Wang2017}. 
The symmetry properties of the target material also have influence on the CEP response of the harmonics. When the interatomic structure is randomly distributed in the crystal on a scale smaller than the laser wavelength, the harmonic spectrum will exhibit $\pi$ periodicity versus the CEP change, because each half-cycle will experience the same average response, like in the case of fused silica \cite{You2017NatComm}. Crystaline quartz possesses a non-centro-symmetric crystal structure, which results in an absence of inversion symmetry, thereby leading to a nonzero nonlinear susceptibility tensor. 
The broken inversion symmetry also ensures that the consecutive laser half-cycles experience a different collective response, leading to 2$\pi$ CEP dependence.

\begin{figure}[!htbp]
    \centering
    % add figure here
    \includegraphics[width=\textwidth]{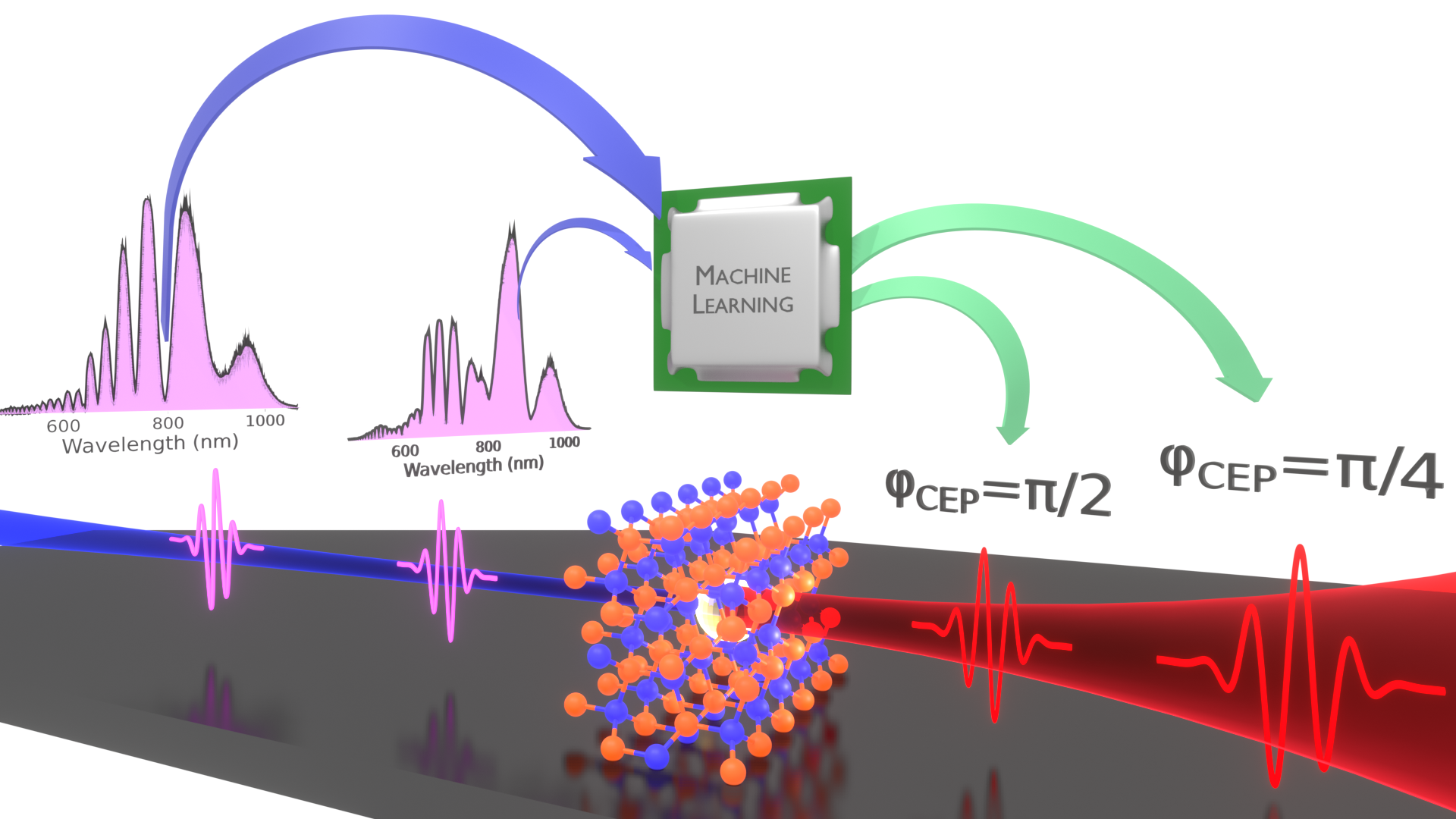}
    \caption{Concept of the experiment. The driving MIR pulses are focused on the ZnO crystal and high order harmonics are generated and detected with a spectrometer. By changing the CEP of the input laser the corresponding harmonic spectrum changes too. A machine learning model is trained to link the laser CEP to the corresponding harmonic spectrum. Applying this model the CEP of the the fundamental laser can be estimated from the observed harmonic spectra. }
    \label{fig:exp-setup}
\end{figure}

By assuring interband contributions to the harmonic generation in a broken inversion symmetry material one paves the way for mapping one-to-one the laser CEP to the generated high harmonic spectrum. This objective is achieved through the utilization of a representative theoretical model for bulk ZnO that incorporates interband \cite{Wang2017} contributions (driving with a MIR laser) to harmonic generation and posseses C6v symmetry group, through a 2-band model.
Then, a machine learning algorithm can be trained with known CEP and solid HHG spectrum pairs to estimate the CEP of a new laser pulse based on the generated harmonic spectrum, as depicted in Figure \ref{fig:exp-setup}.
First, we test and validate this assumption by simulating solid HHG spectra generated from ZnO crystal, applying theoretically constructed laser pulses with known absolute CEP parameter. Subsequently, we demonstrate that the effectiveness of this approach does not depend on an extensive dataset, thereby highlighting its practicality.

In the present work, we select three machine learning algorithms: Linear Regression, Extremely Randomized Trees (ExtraTree) and Gradient Boosting. The linear regressor, the simplest one of all three, works by assuming a linear relationship between the predictor values $X_j$ and the predicted value $Y$, expressed as $Y = \beta_0 + \sum_j \beta_j X_j$.
Major advantages of this algorithm are simplicity and fast training.  The simple dependence on the $\beta_j$ coefficients enables a direct estimation of the significance of certain features. However, the model only performs well if there is a direct linear correspondence between $X_j$ and $Y$. \cite{Altman2015,Krzywinski2015}

The ExtraTree and Gradient Boosting algorithms are both decision tree based ensemble models. In a conventional decision tree model, a tree is formed by segmenting the dataset into branches. The model makes decisions at each node based on feature values, eventually arriving to a prediction at the end-points of the branches. In the ExtraTree model, multiple decision trees are built and used concurrently, and the prediction is decided by averaging the prediction of the independent trees. Different trees are grown from different parts of the training dataset by employing bagging, which decreases the bias and variance in the model compared to a single decision tree. The trees are built by randomly selecting a subset of features in the data, and using random threshold values (as opposed to a certain criteria, such as information gain) to split the nodes of the tree. \cite{Geurts2006}. This high degree of randomization significantly reduces the bias of the model, and makes it less likely to overfit compared to other ensemble models (such as Random Forest)\cite{Ghazwani2023}.

Opposed to the ExtraTree model, where the trees are created by random selections, the Gradient Boosting regression builds the trees iteratively. In each iteration, the model accuracy is estimated by a loss function (typically by the mean squared error, MSE), and new trees are created to correct the error of the previous trees. The final prediction is then given as a weighted average of the prediction of the trees. The weights are updated using a gradient descent method in each iteration of the training. This error-correcting strategy enables higher accuracy, but also can make the model more prone to overfitting \cite{Friedman2001,Ghazwani2023, Natekin2013}.

\section{Simulations} \label{sec:sims}

In order to simulate the CEP-dependent HHG process, we employ an one-dimensional semiclassical model, which provides a computationally
efficient way to compute the response of electrons in a periodic potential to a strong laser field \cite{SzaszkBogr2019}. Herein we adopt the single-electron approximation to derive the Bloch states and their associated energies, enabling us to discern the distinct contributions of various initial states to high harmonic radiation. In more detail, for an electron with charge $e$ and mass $m,$ we consider the following Hamiltonian:
\begin{equation}
H(t)=\frac{1}{2m}(\mathbf{p}-e\mathbf{A}(t))^{2}
+U(x), \label{Ham}
\end{equation}
where velocity gauge is used, $\mathbf{A}(t)$ denotes the time-dependent vector potential component of the external field along the $x$ direction, while $U(x)$ represents the periodic (model) potential of the solid. This simplified model of the potential adequately captures essential features of the crystal lattice's atomic structure, including the band gap width; however, it ignores the three-dimensional symmetries including the broken inversion symmetry of C6v. Consequently, in our investigation of ZnO targets, we appropriately consider factors such as the lattice constant of \SI{5.2}{\angstrom} along c-axis, and a band gap measuring \SI{3.27}{\electronvolt}. On the other hand, scattering events and finite temperature can be taken into account by considering density matrix $\rho$ instead of pure quantum mechanical states (which are practically Bloch waves in this case). That is, we solve the von Neumann equation
\begin{equation}
\frac{\partial}{\partial t}\rho(t)=-\frac{i}{\hbar}\left[H(t),\rho(t)\right] + \left. \frac{\partial}{\partial t}\rho(t)\right|_{scatt},
\label{vN}
\end{equation}
such that the initial density matrix (before the interaction with the laser field) describes thermal equilibrium, and the second term is responsible for the scattering events. We consider the change of the diagonal elements of the density matrix towards thermal equilibrium at a rate of $\gamma_d$, and also the decay of the off-diagonal matrix elements (i.e., the loss of quantum mechanical coherences) with a rate of $\gamma_{od}.$) By considering realistic rates ($\gamma_{od}=\SI{0.25}{\per\femto\second}$ and $\gamma_d=\SI{0.05}{\per\femto\second}$) %\textcolor{blue}{Gergo, please verify.})
, not only the CEP dependence of the harmonic peaks can be determined, but also the most important features of the HHG spectra, namely the presence of the plateau, and the intensity dependence of the cutoff and the heights of the harmonic peaks can be investigated \cite{SzaszkBogr2019}. A numerical solution is obtained using a Cash-Karp Runge-Kutta algorithm. The sinusoidal model periodic potential, $U(x)$ is parametrized by the lattice constant $d$ and the potential depth $U_0$. Corresponding to the eight valence electrons present in one unit cell of ZnO, the value of $U_0$ is chosen so that the band gap between the 4th and the 5th band corresponds to the experimentally determined band gap of ZnO, \SI{3.27}{\electronvolt}, as reported by \cite{Hussain2019}. The band structure of this one-dimensional model is presented on Fig. S4 of the Supplementary document.

\begin{figure}[htbp]
    \centering
    \includegraphics[width=0.85\textwidth]{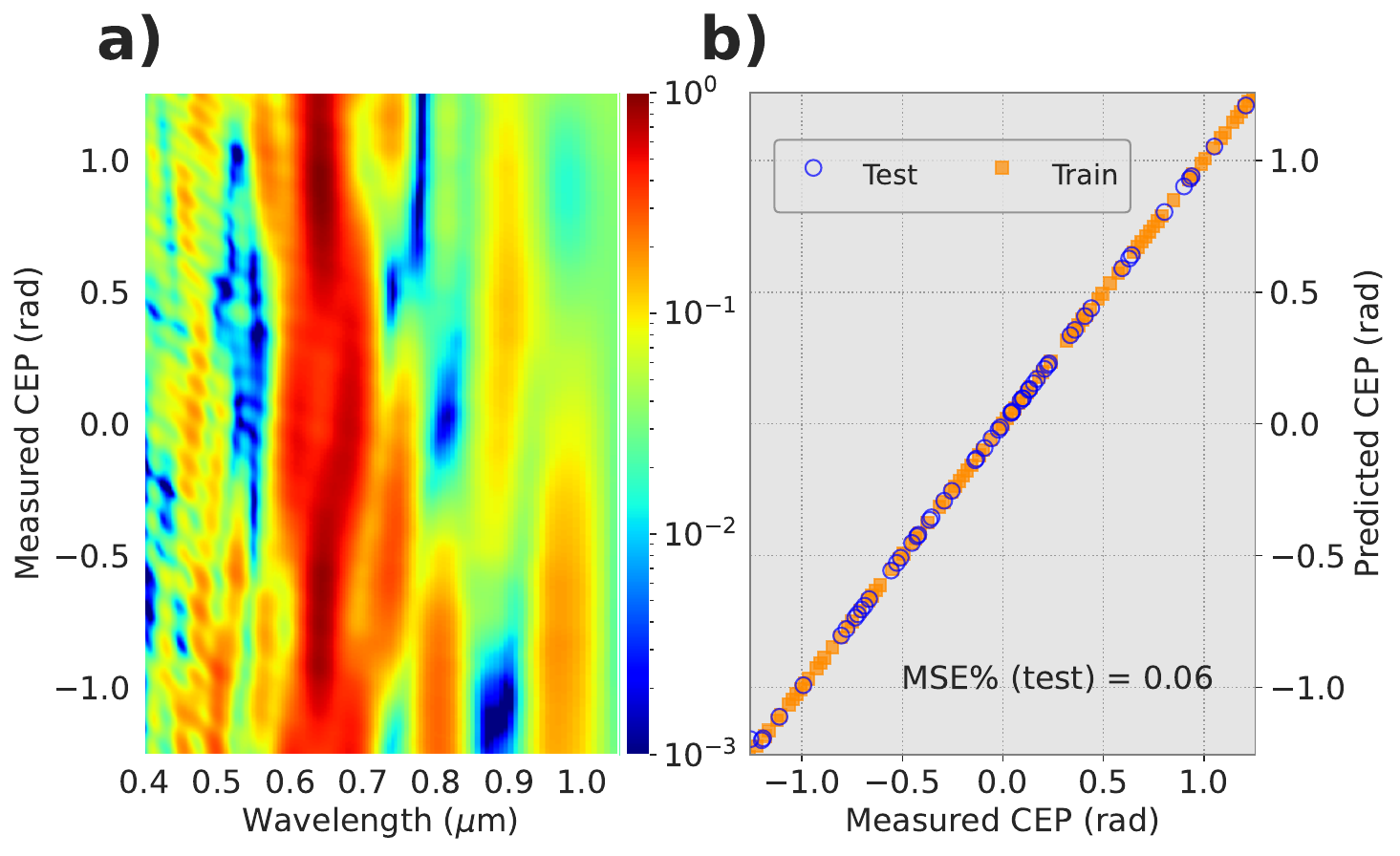}
    \caption{Theoretical CEP scan performed in ZnO with a \SI{18}{\femto\second}, $I_0=\SI{1.36e12}{\watt\per\centi\meter\squared}$, \SI{3.2}{\micro\meter} laser (a). Performance of the machine learning algorithm on estimating the measured relative laser CEP using the train (orange rectangles) and the test (open blue circles) datasets is shown in (b), with respective MSE\% values of 0 and 0.06. The slope of both estimations are  close to 45$^\circ$, representing almost perfect estimating capabilities. One fifth of the test data set is plotted to provide better visual transparency between the train and test data. }
    \label{fig:cep_scan_sim}
\end{figure}

For simulating the HHG process, a laser with $t_p=\SI{18}{\femto\second}$ pulse duration, $\lambda_0=\SI{3.2}{\micro\meter}$ and a peak intensity of $I_0=\SI{1.36e12}{\watt\per\centi\meter\squared}$ (corresponding to a peak field of $E_0 = \SI{3.20}{\giga\volt\per\meter}$) is selected. 
We employ a $\cos^2$ temporal pulse profile as an approximation to a Gaussian profile, leveraging its computational efficiency. Our objective is to model an experiment that can be readily executed within a conventional HHG beamline setup, without necessitating specialized equipment. To this end, the most accessible means for harmonic radiation detection is through the utilization of a spectroscope equipped with a CCD camera. Consequently, we multiply each HHG spectrum with a response function composed of a typical response curve of a grating and a CCD camera, thereby constraining the range of observable harmonics. Details of this response function are reported in the SM.

Using this numerical model, a HHG spectrum is simulated with $\sim$800 different CEP values between $\pm 0.5 \pi$. The result is shown on Fig. \ref{fig:cep_scan_sim}a. 
After generating the simulated CEP-dependent harmonic spectra, they are partitioned into two datasets through random selection. Eighty percent of the data is allocated to the training dataset, while the remaining twenty percent is assigned to the test dataset. The training dataset is employed to train an ExtraTree model, enabling the recognition of patterns linking the spectrum to a specific CEP value. Subsequently, the trained model processes both the spectra from the training and test datasets (the latter representing previously unseen spectra), leveraging the learned patterns to estimate the corresponding laser CEP. As anticipated, the estimated CEP values align perfectly with the training set (particularly considering those data were used to train the model), as shown in Figure \ref{fig:cep_scan_sim}\,b as orange rectangles.

The predictive accuracy of the model is determined by the relative square root of mean square error between the true ($\CEP_t$) and predicted ($\CEP_p$) carrier-envelope phase values expressed in percentages as
\begin{equation}
    \MSE \% = \sqrt{\frac{1}{\Ntest} \sum_{i=1}^{\Ntest} \left(\CEP_{t,i} - \CEP_{p,i}\right)^2} \cdot \frac{100}{\pi}.
\end{equation}

When applying the model on the test set (blue circles), very good agreement is achieved between the true and the estimated CEP values, with an MSE\% of 0.06. The outcome proves our assumption, that there is a one-to-one link between the harmonic spectrum and the CEP and this can be captured with a machine learning model.
The excellent agreement observed between the true and estimated CEP values can be attributed in part to the extensive sampling of training data. Nevertheless, acquiring a sufficient number of experimental data points may pose challenges in practical scenarios. Generally, increased training data enhances model accuracy. Therefore, it is important to discuss the effect of the number of training data ($\Ntrain$) on the model accuracy (MSE\%) using the test data. Furthermore, in a general case, the data used for model training is picked on randomly selected label values. This has the disadvantage of an uneven sampling of the parameter space, which in turn results in a varying model accuracy over the parameter range. However, in an experimental situation where the tag values are known to be within a closed range, it is possible to provide the model with an evenly sampled training data.

\begin{figure}[htbp]
    \centering
    % add figure here
    \includegraphics[width=0.8\textwidth]{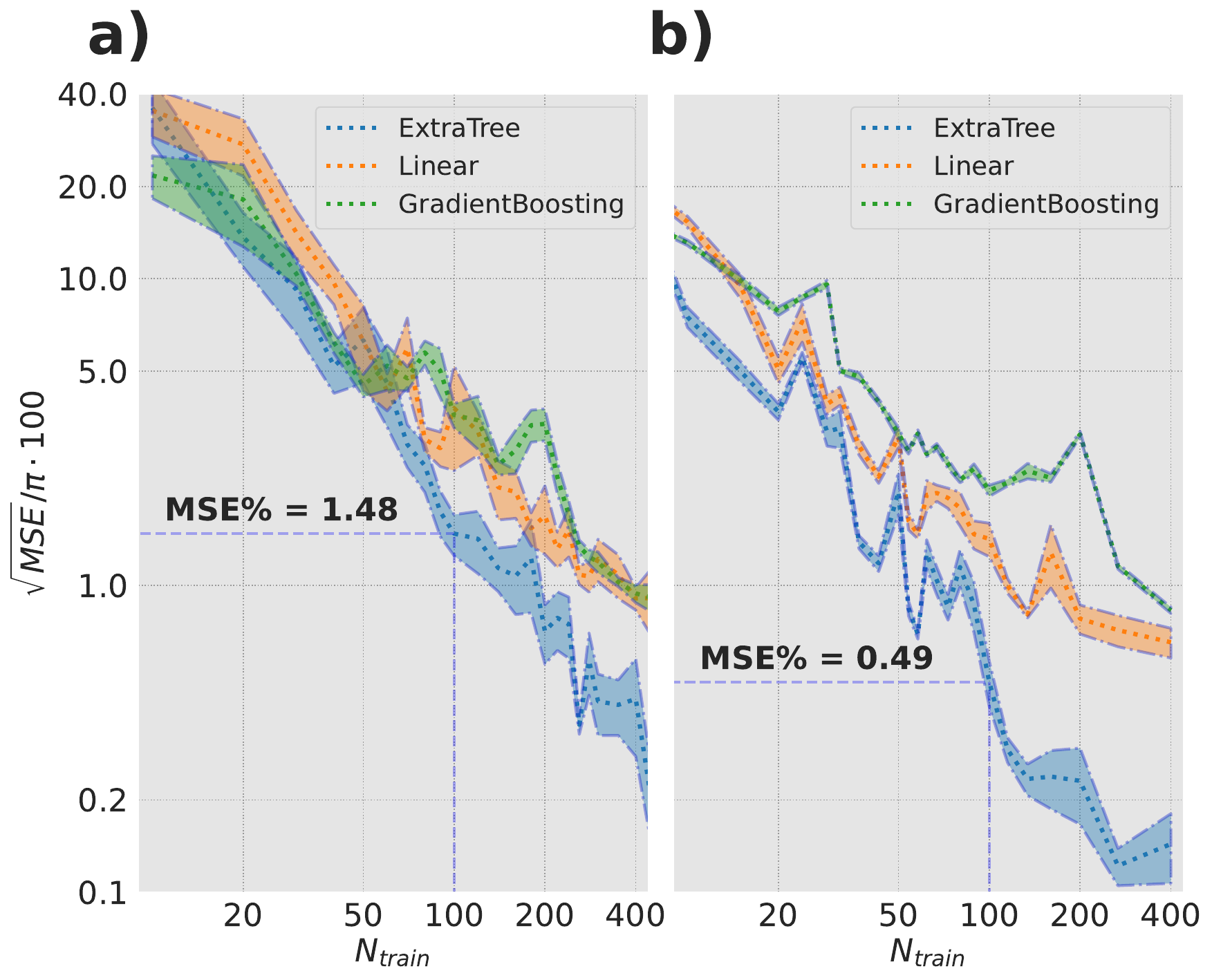}
    \caption{Comparative analysis of model predictive accuracy between a) random and b) equidistant CEP sampling of the training data. The precision is given by the MSE\% of the linear fit on the prediction from 150 test datasets at randomly picked novel CEP values. The random selection is done 10 times at each $N_\mathrm{train}$ value allowing to give error estimates to the predictions (shaded area on the plots). In both randomly and evenly sampled CEP scenarios, the model performance with the test data is given on the plots for $\Ntrain = 100$ for the ExtraTree model. 
    }
    \label{fig:ntrain-effect}
\end{figure}

To address this question, we train three different models (ExtraTree, Linear Regression and GradiantBoosting) using a varying number of train data ($\Ntrain$) and a fixed number of test data ($\Ntest$). We have simulated a total of 800 solid HHG spectra corresponding to an input of an equidistant CEP grid between $-0.4\pi$ and $0.4\pi$ (in the simulation, because of the 1D model, the CEP has $\pi$ periodicity, therefore we reduced the examined CEP range, to keep one-to-one mapping). We used random sampling without replacement to select $\Ntrain=1 - 400$ pairs and for each $\Ntrain$ case we selected in a similar manner $\Ntest = 150$ and calculated the model's performance, shown in Figure \ref{fig:ntrain-effect}\,a.
% For each case, we randomly select $N_{selected} = N_{train} + N_{test}$ data without repetition out of a total of $N_{total} = 800$ simulated HHG spectras, which are calculated on an equidistant CEP grid between $-0.4\pi$ and $0.4\pi$.
This random selection is performed 10 times for each $\Ntrain$ values. The model is trained using the selected $\Ntrain$ HHG spectra, and then tested using the $\Ntest$ data points to estimate the CEP from spectra that was not known for the model before. The $\MSEper$ quality parameter is then obtained by performing a linear fit on the predicted versus actual CEP values of the test dataset.

Figure \ref{fig:ntrain-effect} presents a comparative result of this analysis of the effect of $\Ntrain$ using randomly (a) and equidistantly (b) picked CEP training data. Regarding the models, we can observe similar trends in both cases. The ExtraTree model presents superior model accuracy in case of low number of training data ($\Ntrain < 100$), which can be attributed to the algorithm's marked resilience againist overfitting. However, the linear model, which is the least accurate in this sparsely sampled regime, achieves superior accuracy in case of a densely sampled training data ($\Ntrain > 200$). This hints to the existence of a strong linear correlation between the CEP of the laser and the features of the HHG spectra. The GradientBoosting method stays in-between for small train sets, and performs comparatively poorly when a high number of training data is available. As expected, in both sampling cases, at low number of training data the performance improves exponentially (note the log-log plot) with the number of the training data, but around $\Ntrain=200$ the prediction accuracy starts to saturate, indicating that additional training points will not improve substantially the model's precision.

It is also observed that the equidistant sampling (Figure \ref{fig:ntrain-effect}\,b) of the training data results in a significantly improved accuracy in all cases, 
resulting in an increase of the accuracy by about an order of magnitude, especially in the case of low ($\Ntrain < 200$). This implies that the number of required training data points for a given accuracy can be significantly reduced by performing a methodical CEP sampling in the experimental case. The improvement is especially prominent in case of the ExtraTree model since it gains an order of magnitude prediction performance in favor of the evenly sampled scenario.

\section{Experimental setup}

\begin{figure}[htbp]
    \centering
    \includegraphics[width=0.9\textwidth]{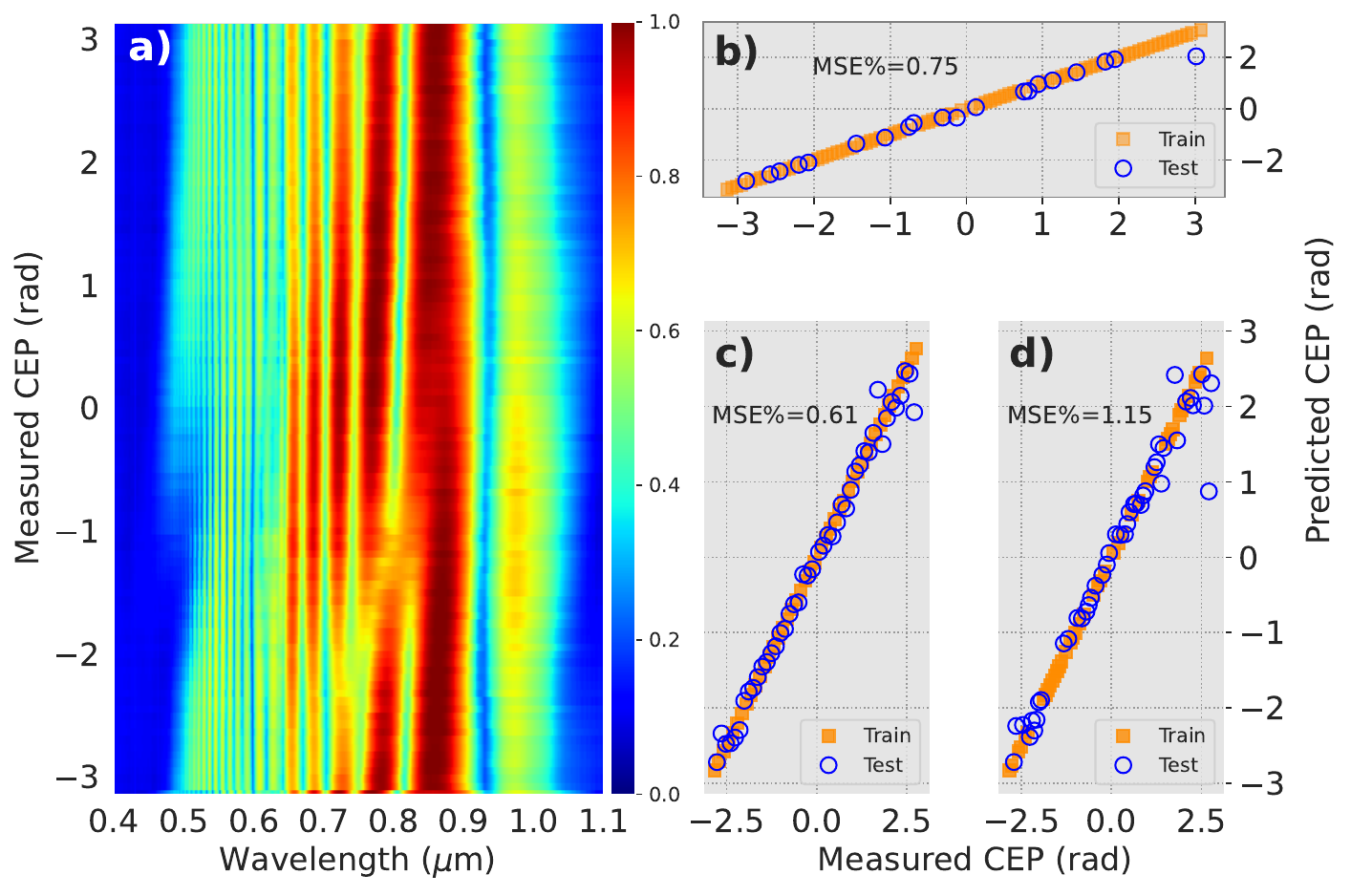}
    \caption{Experimental CEP scan in ZnO performed with a \SI{18}{\femto\second}, \SI{0.13}{\watt}, \SI{3.2}{\micro\meter} laser (a). The performance of the ExtraTree machine learning algorithm on estimating the measured relative laser CEP using (b) 80\% train (orange rectangles), 20\% test (open blue circles); (c) evenly while (d) randomly sampled 50\% train and 50\% test datasets. The slope of each prediction scenario lies very close to 45$^\circ$, representing almost perfect estimating capabilities, with MSE\% values on the test sets: (b) 0.75, (c) 0.61, (d) 1.15. }
    \label{fig:exp-vs-sim}
\end{figure}

The experiments were carried out using the MIR laser at ELI-ALPS \cite{Thire2018}. The laser is operating at 100 kHz repetition at \SI{3.2}{\micro\meter} and capable to deliver \SI{140}{\micro J} of pulse energy. The measured spectrum of the driving field can be seen in Figure S1 in the Supplementary document. The 45 fs output pulses are spectrally broadened in BaF2 and Si optical windows and recompressed in bulk BaF2 windows combined with three reflections on negative TOD dispersive mirrors to reach 18 fs pulse duration.
The temporal profile of the pulse was measured by a TIPTOE device by coupling out the beam before the off-axis parabolic mirror.
The accurate CEP measurement and control is crucial for this experiment, we used an f-2f setup called Fringeezz by Fastlite which is controlling the acousto-optic programmable dispersive filter (AOPDF, Dazzler, Fastlite) in the OPCPA front-end in a closed loop. The device is able to measure CEP values shot-to-shot at 10 kHz, while the laser is capable to deliver 100 mrad CEP stability. We generate high harmonics in a \SI{90}{\micro\meter} thick ZnO crystal by focusing \SI{1.3}{\micro J} pulses using an off axis parabola with \SI{100}{\milli\meter} focal length. The resulting intensity for the compressed pulses was 1.3$\cdot$10$^{12}$ W/cm$^2$, by introducing chirp we lowered this intensity to 6.6$\cdot$10$^{11}$ W/cm$^2$. After solid HHG, a thick \SI{50}{\milli\meter} long BK7 bulk filters out the MIR driving beam and transmits the harmonics, but cuts off radiation below \SI{350}{\nano\meter}. Subsequently, the remaining visible range of the harmonic spectrum is imaged into a commercial spectrometer (AvaSpec-ULS2048 from Avantes). In this configuration we were able to measure from harmonic order 3 to harmonic order 9. During the measurement we recorded 100 spectra for each CEP settings and used the average of these during the analysis, as shown in Figure \ref{fig:exp-vs-sim}\,a. The CEP scan was recorded between -$\pi$ and $\pi$ having 100 evenly distributed CEP values in between. The spectrum for each CEP value is normalized in this plot.

The fringe pattern extending over the NIR region on the harmonic spectrum on Figure \ref{fig:exp-vs-sim}\ is solely due to interference of harmonics generated from the front and the rear side of the ZnO crystal \cite{Hollinger2020} and is therefore independent of the CEP. However, the harmonic spectrum expresses harmonic minima at different CEP values, which is the result of constructive and destructive interference between the XUV bursts generated in consecutive laser half cycles \cite{You2017, You2017NatComm}. 

Direct comparison between simulation and experimental data is not straightforward, because on one hand, in simulation the CEP is expressed in absolute terms, while in experiment it is relative. On the other hand, the simulation does not take into account the complexity of the experimental conditions, and the 1D model cannot be expected to reproduce neither 3D symmetry properties of the real crystal, nor propagation-related effects, like phase matching.  However, the spectral minima shift as the function of CEP is visible in both simulations and experiments. It is known that the extent of CEP sensitivity has dependence on the instantaneous structural changes in the laser-induced lattice, which is not taken into our simulations, thereby possibly causing some difference in features. Furthermore, it is important to note that the CEP scan confirms a one-to-one mapping between the CEP and the related spectra, as required for the machine learning model to work. 

We split the experimentally recorded data in Figure \ref{fig:exp-vs-sim}\,b into train (80\%) and test (20\%) randomly selecting into these datasets and present the performance of a trained ExtraTree model in estimating the laser's CEP from the input spectra. The orange rectangles represent the CEP predictions using the spectra from the train data set, showing perfect agreement between the predicted and the true laser CEP. The blue circles present the performance of the model on the previously unseen test data, indicating that the model is well trained to recognize spectral patterns in the solid HHG spectrum to estimate the laser CEP correctly with an MSE\% = 0.75. The only major discrepancy between the estimated and the true CEP is where there is a gap in the sampling in the train data set (around $\pi$\,rad in the figure). This issue can be overcome by evenly sampling the training dataset, as described in Figure \ref{fig:ntrain-effect}\,b. To visualize this effect we resample our dataset into 50\% train and 50\% test with evenly and randomly sampled input CEP values, as shown in Figure \ref{fig:exp-vs-sim}\,c and d, respectively. As expected, the evenly sampled scenario (MSE\% = 0.61) is outperforming the randomly sampled case (MSE\% = 1.15) by a factor of 2 in estimating the CEP of the laser from the harmonic spectra. 
These outcomes clearly prove that our concept for CEP estimation works and could be used for monitoring laser CEP during experiments. They also highlight that good model performance can be only achieved if the training CEP values are evenly sampled with small step size.

\section{Conclusion and outlook}

In conclusion, we have laid out a conceptual scheme for estimating MIR laser CEPs relying on the spectrum of high order harmonics generated from a solid crystal exploiting the ability of complex pattern recognition of a machine learning model. Furthermore, we demonstrated the applicability of this concept both through theoretical simulations and experimental measurements. This proven scheme offers an economic, instrumentally not demanding option to measure the laser CEP. The concept can be generalized to other laser wavelength, assuming that the combination of the laser and the crystal fulfills the requirements of one-to-one mapping of the CEP to the solid HHG spectrum and the existence of 2$\pi$ periodicity. Furthermore, in principle it is possible to perform single shot recordings of the harmonic spectrum with a sampled beam (only 1\% of the total energy was used) while an experiment is carried out with the remaining beam.  In case of random CEP laser source, the latter approach will allow CEP tagging.

The current study relies on a reference method as implimented in a fast CEP measurement device, Fringeezz by FastLight, to train the ML model. However with more accurate modelling the simulated and experimental CEP scans could match better, meaning simulations could be a tool to train the ML model, as it was done in \cite{Toth2023} for FROG scans. Additional benefit of such simulation would be to retrieve the absolute CEP of the laser, since in simulations it is an input parameter.
It is important to notice that this study implies a fixed laser intensity for the training and for subsequent actual measurements and CEP retrieval. The question arises then whether intensity variations would ruin the process. Initial investigations have been carried out that show the ability of the machine learning to recover CEP values from various field configurations and intensities. However, in-depth study of the sensitivity to intensity fluctuations is ongoing. 

Overall, our proposed protocol, combined with precise simulations, opens up the route towards multi-parameter estimation and optimization using machine learning concepts, as it has been already proposed by \cite{Sanchez-Gonzalez2017, Klimkin2023, Yang2022}. Solid HHG is a process, where the laser parameters and the material features have a direct manifestation in the measured spectrum, resulting in an economic and adaptable tool for monitoring experimental conditions. Therefore we utilize HHG signal from semiconducting ZnO crystal in MIR regime to infer the CEP, based on ML models trained with combined experimental and simulations results.
Our findings demonstrate the effectiveness of the ExtraTree-based models for predicting the one-to-one correlation between CEP and HHG spectra.
Our approach can prove instrumental in determination and control of laser parameters, such as intensity, CEP in-situ depending on the target material, prior to an advanced HHG based experiment design. While it is possible to infer CEP from spectral measurement, this usually needs extended data acquisition. Our method achieves the same with significantly less number of measurements, thereby being of prominent use towards reducing cost of running experiment campaign, as relevant for large-scale user facilities like ELI ALPS.

\section{Funding}

The ELI-ALPS project (GINOP-2.3.6-15-2015-00001) is supported by the European Union and co-financed by the European Regional Development Fund.

\section{Acknowledgments}

We thank Rajaram Shrestha for his kind help in processing data and plotting figures for the Supplementary Material.

%%%%%%%%%% If using BibTeX:
\bibliography{references}

\end{document}